# Defect related photoluminescence and EPR study of sintered polycrystalline ZnO

Sanjiv Kumar Tiwari[*]
Department of Physics, Indian Institute of Technology Kanpur, Kanpur- 208016, India.
Department of chemical sciences, Tata Institute of Fundamental Research Colaba Mumbai

**ABSTRACT**

We report on low temperature photoluminescence (PL) and EPR study of bulk polycrystalline ZnO . Variation of peak position of donor bound exciton transition ($D_1X_A$) and $FX_A^{n=1}$-1LO with pump intensity show red shift and blue shift respectively due to exciton-exciton scattering. EPR spectra reveals three peaks at g value of 1.985 , 1.956 and 1.939 respectively, g=1.956 and g=1.985 is due to shallow donors Zn interstitial and oxygen vacancy respectively An EPR spectrum at 100 K reveals higher degree of asymmetry and hyperfine splitting due to crystal field and inhomogeneous relaxation of paramagnetic centers. Strength of crystal field splitting (CFS) and spin orbit coupling (SOC) in sample is discussed using PL peak position of various excitonic emissions. Whereas, inhomogeneous relaxation of paramagnetic centers is discussed in terms of their activation energy during thermal quenching process.

## I. INTRODUCTION

Zinc oxide is a semi-conducting material used in variety of applications, ranging from sunscreen lotion and talcum powder to piezoelectric transducers and phosphors.[1,2] The band gap structure and optical properties of ZnO are similar to GaN, however, it has advantage over GaN in terms of strong binding energy of excitons. Higher binding energy of exciton (60 meV) in ZnO makes it a potential candidate for optoelectronic devices and ultra violet LEDs.[3,4] The low temperature PL profile of ZnO single crystal has been investigated by several researchers,[5-10]. It has been widely reported that PL profile of ZnO at low temperature comprises of several bound and free exciton transitions and a dominating line is due to neutral bound exciton. The peak position of neutral bound exciton varies from sample to sample, for example, the position of most intense line in PL profile of ZnO has been reported from 3.3624 to 3.3628 eV by different groups.[7,11-16] ZnO is n-type material with dominant donor being either a native defect such as oxygen vacancy ($V_0$) and/or the zinc interstitial ($Zn_i$).[17] Krooger assigned $V_0$, zinc vacancy ($V_{Zn}$) as dominant donor and acceptor respectively.[18] It has been concluded that neural oxygen vacancies are considered to be a major component of the defect structure of ZnO. EPR and first principal calculations show that $V_0$ is deep donor.[19-21] Hall measurement on electron irradiated ZnO have provided supporting evidence that zinc interstitials may be a residual shallow donor.[22] In this paper, we report low temperature PL study in temperature range of 6K-200K and EPR measurement in temperature range of 275-100 K of sintered bulk polycrystalline ZnO. Since we used pulsed laser as an excitation source, hence, dynamics of donor bound exciton with excitation intensity (local heating) and with temperature (bulk heating) is also discussed.

## II. EXPERIMENTAL DETAILS

ZnO samples were prepared by cold press of ZnO powder 99.999% (Sigma-Aldrich) at pressure of 6 Ton followed by sintering at 1000 $^0$C in air for five hours. In order to see the effect of sintering on crystal structure and stress, X-ray diffraction (XRD) spectra of the sample were taken before and after sintering using CuK$_\alpha$ source ($\lambda$=1.54 A$^0$). Steady-state PL measurement was carried out with sintered sample placed in a closed cycle cryostat (ARS, Model 830) in the temperature range of 6-250 K. Third harmonic of Nd:YAG ( DCR-4G, Spectra Physics $\lambda$=355 nm, 4 nano-second pulse duration, 10 Hz repetition rate, 2 mJ/pulse energy ) was used as excitation source. Laser was focused on the sample with help of lens; excitation area was circular of diameter 400 μm. The emission light was collected by a lens, imaged onto fiber-coupled monochromator ( Sahmrock SR 303i, Andor Technology) and detected by intensified charge couple device (ICCD DH720, Andor Technology). ICCD was operated in gating mode and emission was collected for one microsecond ( gate width one micro-second) after laser pulse ( gate delay zero) because diffusion of excited exciton out of the focal volume/excited volume can be neglected at this time scale. Spectral resolution of spectrograph was ascertain using grating property and active area of ICCD. Grating of 1200 lines/mm, 1.66 nm/mm dispersion and 750 pixel (active area of ICCD, 1 pixel=25 micron) gives observation window of 31 nm. Further, calibration with hollow cathode neon lamp provides dispersion of 0.05 nm/pixel. PL peak position was determined with respect to excitation wavelength with accuracies of ± 1meV. The excited exciton density $n_{ex}$ within the sample by UV illumination is calculated using rate equation $n_{ex}= \frac{\eta\lambda}{dhc}\tau_{Fx}p$. Where $P$ is the excitation intensity, $\lambda$, $h$, $c$ is the excitation wavelength, plank's constant, velocity of light in vacuum respectively . $d$ ( $\approx$ 500 nm) is the thickness of active layer, $\tau_{Fx}$ ($\approx$ 1x 10$^{-9}$ sec ) is radiative life time of free excitons and $\eta$ is coupling efficiency .[23,24] $\eta$ is deduced by assuming 14% of light is reflected from the sample and quartz window and most 50% of excitation intensity is converted to create electron-hole pair. This leads to the exciton density of 10$^{18}$-10$^{20}$ cm$^{-3}$ within focal volume with increase of excitation intensity from 9.10 kW-cm$^{-2}$ to 3.87 MW-cm$^{-2}$. Further assuming excitation area as cylinder of length α$^{-1}$ (α=2 x 10$^5$ cm$^{-1}$ at 355 nm) and diameter 400 micron and exciton density $\approx$ 10$^{20}$ cm$^{-3}$ gives average separation between excitons (<r$_{ex}$>) of about 100 A$^0$ .[25] This implies that <r$_{ex}$> is almost three times larger than exciton diameter (35 A$^0$). These calculations also agree with Klingshrin, who propose that maximum generation rate of exciton with excitation power of 500 kW-cm$^{-2}$ in fundamental edge with UV illumination can be taken as 10$^{29}$ cm$^2$-sec$^{-1}$. This assumption also gives exciton density of order of 10$^{20}$ cm$^{-3}$.[26, 27] These calculations suggest that exciton-exciton interaction is very strong in our case. Whereas, formation of new complex like bi-exciton or exciton molecules is less likely and formation of electron –hole plasma (EHP) is negligible. The EPR experiment was performed in temperature range of 275-100K, with microwave frequency 9.412 GHz, and 2mW power (Bruker BioSpin). The g value was calculated using DPPH (2,2-Diphenyl, 1-picryl hydrazyl, g= 2.0036) as reference. For EPR measurements sintered ZnO powder was placed in a glass tube (~15 cm long, 6 mm inner diameter, 8 mm outer diameter). Sample tube was evacuated to $10^{-5}$ torr, and a flame from torch was used to remove contamination from the surface. The torch was then used to seal the tube while maintaining the vacuum inside.

## III. RESULTS AND DISCUSSION

Figure 1 shows, XRD spectra of sintered and un-sintered ZnO sample, solid line and dotted line are XRD peak of un-sintered and sintered ZnO powder respectively. XRD spectra reveals that sample is polycrystalline in nature having multi diffraction peaks corresponding to [100], [002], [101], [102], [110] and [103] plane. XRD spectra of sintered ZnO get shifted towards higher angle and FWHM (full width at half maxima) of peaks decreases. Average grain size is calculated by using FWHM of all diffraction peaks. Grain size is found to be 19nm and 28 nm, before and after sintering respectively. Increase of grain size after sintering indicates that grain growth is driven by the grain boundary surface tension. Decrease/increase of FWHM/grain size of diffraction peak after sintering is in consistence with earlier reported results.[2]

Width of XRD peaks is due to grain size ($B_{size}$), strain ($B_{strain}$) and instrumental broadening ($B_{inst}$). Therefore, total width can be written as.[28]

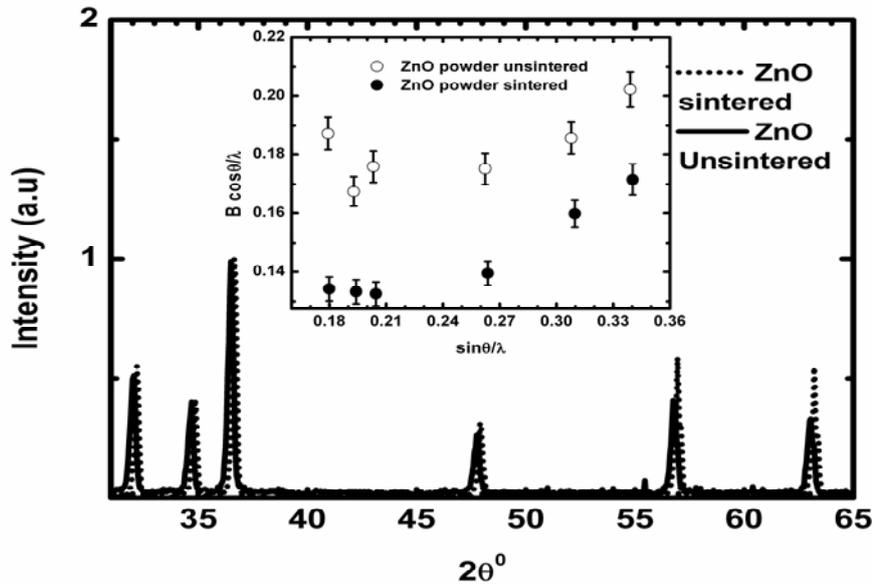

**Figure 1**. XRD spectra of sintered and un-sintered ZnO sample, inset shows Wilson-Hal plot of sintered and un-sintered ZnO sample.

$$B_{total}(cos\theta)/\lambda = 0.9/t + 2\frac{\Delta d}{d}\sin\theta/\lambda + B_{inst}\cos\theta/\lambda \qquad 1$$

Where, first term is Debye-Scherrer term, second term is strain term and third term is instrumental broadening. Slope of plot of $B_{total}\cos\theta/\lambda$ Vs $\sin\theta/\lambda$ as shown in inset of figure 1 gives strain. Since, $\Delta d/d$ is reduced on sintering inhomogeneous strain can be ruled out whereas, homogeneous strain may be present in sample because of observed asymmetry of XRD line as shown in figure 1.

Figure 2 shows the PL profile at 6K, the PL peaks position is determined by averaging the PL profile over ten data points. In figure 2(a) the free excitonic transition is observed due to energy state n=1($FX_A^{n=1}$) and first excited energy state n=2 ($FX_A^{n=2}$) at 3.378 eV and 3.414 eV respectively close to the predicted value of A exciton. [28, 29] Following the reported energy separation of $FX_A^{n=1}$ and $FX_B^{n=1}$, we have assigned the emission peak at 3.391 eV as $FX_B^{n=1}$,. The observed energy separation of $FX_A^{n=1}$ and $FX_B^{n=1}$ is 13 meV, close to predicted experimental value of 9-15 meV.[15, 30, 31] Transition energy of $FX_A^{n=2}$ is less then earlier reported value of 3.422 eV. This is due to polycrystalline nature of sample. We did not observe any signature of C exciton because it is expected that C excitons are thermalised down to lower exciton level at low temperature. At 6K most intense transition is due to donor bound exciton $D_1X_A$ appearing at 3.347 eV. This line corresponds to A-excitons bound to shallow donor with spectral width (FWHM) of 15 meV. Due to large spectral width, $D_1X_A$ seems to be envelop of two peaks at 3.347 and 3.354 eV respectively. Thus, it is expected that A exciton is bound to two different shallow donors.

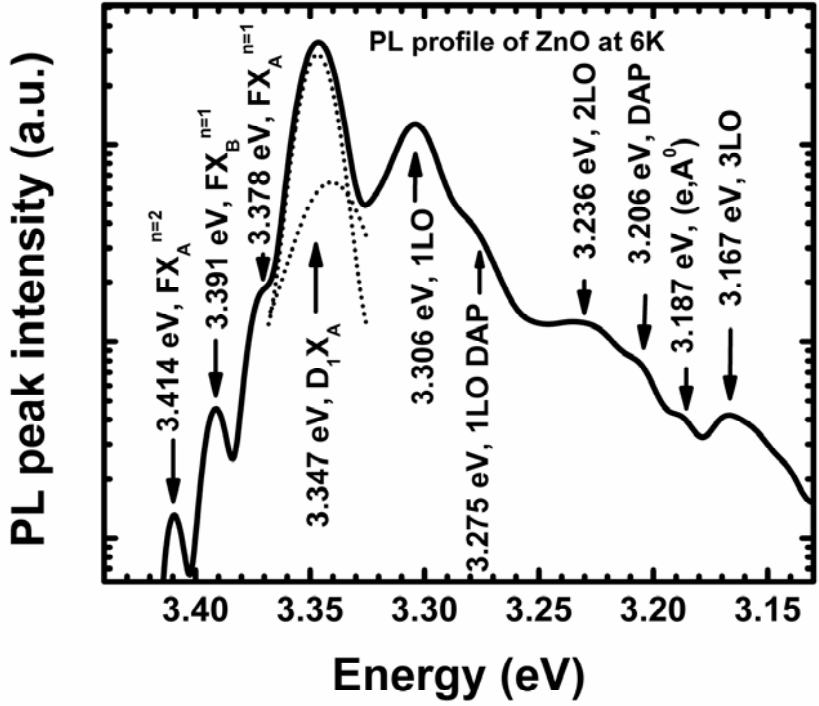

**Figure 2**. Photoluminescence profiles of bulk ZnO at excitation intensity of 33 kW-cm$^{-2}$ dotted line is Gaussian de-convolution of D$_1$X$_A$ peak at 3.347 eV and 3.354 eV.

Presence of two different donors was verified by EPR experiment and discussed in next section. The other peaks observed on lower energy side of main PL profile are assigned as LO replicas of FX$_A^{n=1}$. Phonon replica occurs due to free exciton transition as well as due to bound exciton transition. First (1LO), second (2LO) and third (3LO) longitudinal optical phonon replica of FX$_A^{n=1}$ appears as a shoulder at 3.306, 3.236 and 3.167 eV respectively in main PL profile, whereas, 1LO phonon replica of D$_1$X$_A$ appear at 3.275 eV. Energy separation between FX$_A^{n=1}$ and its phonon replicas (m LO, m=1,2,3) are 72, 142 and 211 meV respectively. Following Hall measurement data at low temperature transition at 3.187eV assigned as free electron acceptor peak (e, A$^0$). Energy difference between FX$^{n=1}_A$ and (e, A$^0$) transition is 191 eV close to the reported value of 195 meV.[8, 11] In order to confirm the dominance of bound exciton transition at low temperature. Variation of PL peak position and PL peak intensity with temperature is shown in figure 3.

Variation of PL peak position of D$_1$X$_A$ versus temperature is shown as solid points and fitted with Varshni's formula,[32]

$$E_g(T) = E_g(0) - \frac{\gamma T^2}{(\beta + T)} \qquad 2$$

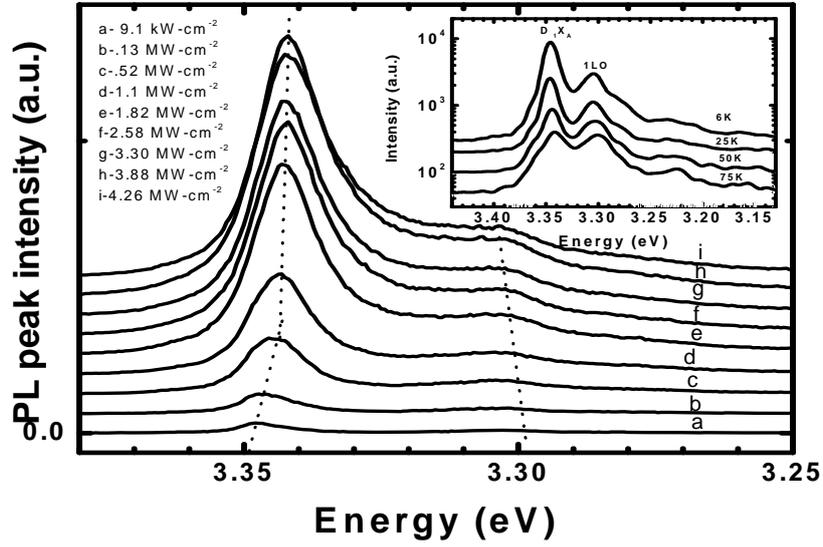

(a)

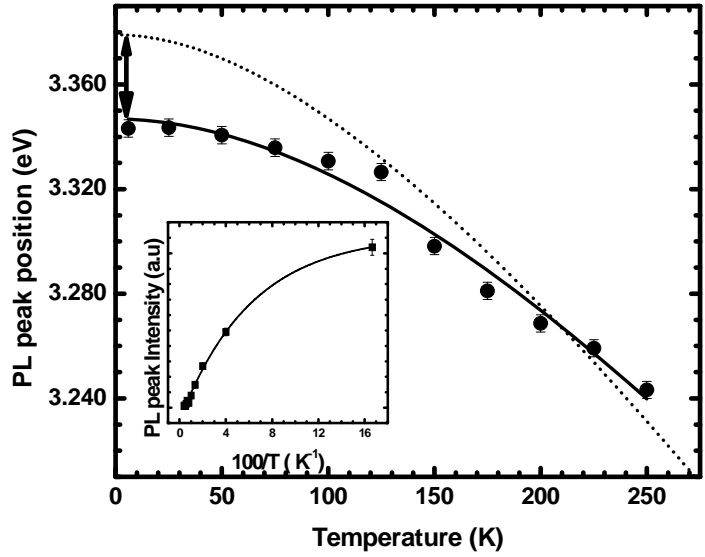

(b)

**Figure 3**. Variation of PL peak position with temperature at excitation intensity of 33 kW-cm$^{-2}$, inset shows variation of PL peak intensity with temperature. (b) Variation of PL profile with temperature at excitation intensity of 4.5 MW-cm$^{-2}$, inset shows temperature variation at excitation intensity of 0.5 MW-cm$^{-2}$. Solid line is theoretical fit of equation 2 to experimental data points.

Where T is the absolute temperature, $E_g(0)$ is the band gap energy at T=0, and $\gamma$ and $\beta$ are constants. The $\gamma$ and $\beta$ value obtained from fitting is 7.2 x $10^{-4}$ eV/K and 105 K respectively. Solid line indicates theoretical fit to equation 2, where as dotted line is function fit with same constant and $E_g(0)$ = 3.391 eV. PL peak position shifts towards higher energy side with decrease in temperature. This behavior of sample is due to lattice contraction at lower temperatures. Using x-ray powder diffraction it has been shown that lattice parameter $a$ decrease from 3.250 $A^0$ to 3.247 $A^0$ with decrease in temperature from 296 to 4 K.[33] We also estimated the exciton donor binding energy from the difference between the bound exciton (D $_1X_A$) energy and the free exciton energy. The exciton donor binding energy is roughly 24 meV. The variation of PL peak intensity versus temperature is shown in inset of figure 3 and fitted using formula

$$I(T) = \frac{A_1}{1 + C_1 \exp(-E_a / k_B T)} \qquad 3$$

Where $E_a$ is the activation energy in the thermal quenching process, $A_1$, $C_1$ are temperature independent constants and $k_B$ is Boltzmann constant. Solid line shows theoretical fit to experimental data points. From intensity plot, activation energy is found to be 12 meV for bound exciton transition, implying dominance of donor bound exciton below 144K.

Inset of figure 3 show variation of PL profile with temperature at constant excitation intensity 0.5 MW-$cm^{-2}$. As temperature increases, either by local heating (by laser radiation) or bulk heating (temperature variation of sample) the peak of D $_1X_A$ and $FX^{n=1}_A$ -1LO get broader and finally merge into a single broad peak.

Since the peaks are well resolved at 6K, thus PL at 6 K was chosen to study the dynamics of D $_1X_A$, $FX^{n=1}_A$ -1LO and DAP with increase of excitation intensity. Figure 4(a) shows the variation of peak position of D $_1X_A$ and $FX^{n=1}_A$ -1LO with increase of excitation intensity. The peak position of D $_1X_A$ is red shifted while $FX^{n=1}_A$ -1LO peak is blue shifted and finally both settled at constant value. The PL peak energy is given as

$$\hbar\omega_{PL} = E_{FX} - E_b^{FX}(1 - \frac{1}{n^2}) - \frac{3}{2}kT, n > 1.$$

Where, $E_{FX}$ is energy of free exciton, $E_b^{FX}$ is binding energy of free exciton (60 meV) and 3/2 kT is thermal/kinetic energy of excitons. Assuming negligible kinetic energy of bound excitons comparable to free exciton, peak energy for bound excitons can be written as

$$\hbar\omega_{BX} = E_{BX} - E_b^{BX}(1 - \frac{1}{n^2})$$, hence $D_1X_A$

is red shifted by 9 meV. This is in agreement with experimental result. Above said statement is mostly valid for exciton –exciton interaction processes. Inorder to show that exciton-exciton interaction is dominate process in our case figure 4(b) shows the shift of D $_1X_A$ line ($\Delta E$) with increase of excitation intensity (P) at different temperature. It shows power law behavior as $\Delta E \approx P^{0.90\pm.03}$ at 6K, whereas other process like exciton-electron interaction and EHP interaction shows power law as $\Delta E \approx P^{2/3}$.[26] Therefore, red shift of D $_1X_A$ is due to exciton-exciton scattering.[34].

A general relationship for the emission lines involving phonon and exciton emission can be written as $E_n = E_0 - (n\hbar\omega_{LO} - \Delta E')$, where $\hbar\omega_{LO} = 72$ meV, $E_0$ is exciton energy and $\Delta E' = \frac{\hbar^2 k^2}{2M}$ is kinetic energy of free exciton and is equivalent to $\frac{1}{2}k_B T$.[35] However, the energy separation between the $FX_A^{n=1}$ and $FX_A^{n=1}$-1LO emission peak exhibit a strong temperature dependence. The spectral line shape $I(LO)$ of $FX_A^{n=1} - 1LO$ has the form $I(LO) \approx E^{1/2} \exp(-\frac{E}{k_B T}) P(E)$ where E equals $\hbar\omega - (E_{FX_A^{n=1}} - \hbar\omega_{LO})$ and $P(E)$ is the transition probability which varies as $E^m$ for the corresponding phonon assisted transition, where m is constant. At low temperatures (< 75K) P(E) is typically assumed to be proportional to E.[36] Therefore, the peak position of 1LO band will be shifted by $3/2k_B T$ to the higher energy side. In our case, fig. 4(a), $FX_A^{n=1}$-1LO peak gets blue shifted by 3 meV.

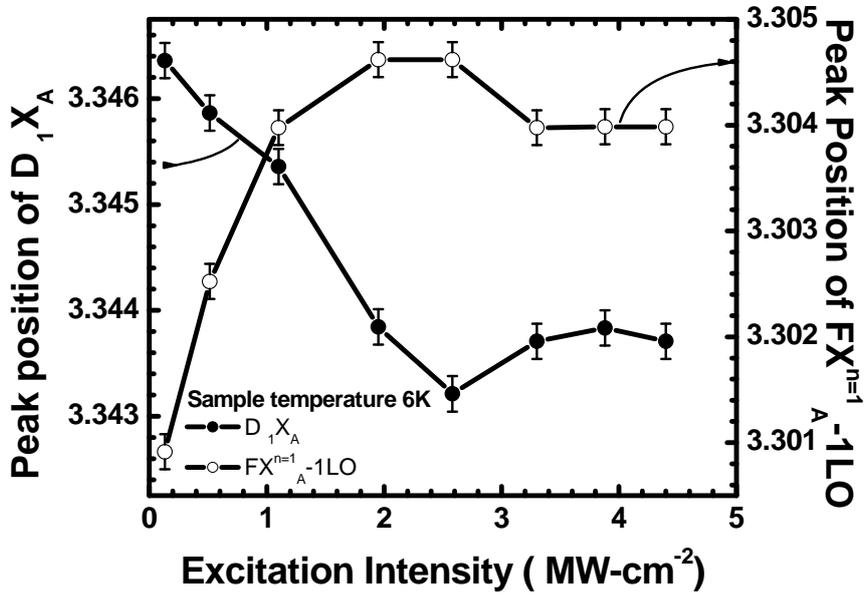

(a)

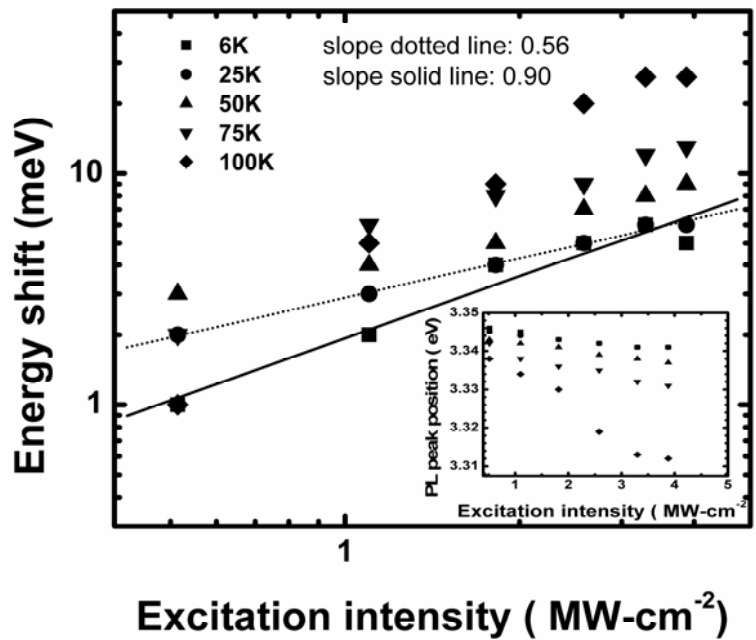

(b)

**Figure 4**. Variation of PL peak intensity with pump intensity at fixed 6K, (a) variation of PL peak position of $D_1X_A$ and FX-1LO with excitation intensity at 6K. (b) Variation of energy shift with excitation intensity at various temperature, inset shows variation of PL peak position with excitation intensity at various temperature.

To understand the behavior of DAP peak with intensity-induced temperature, consider the DAP peak position given by,[37]

$$\hbar\omega = E_g - \left\{(E_D + E_A - \frac{e^2}{\varepsilon r_{DA}} - (\frac{e^2}{\varepsilon})(\frac{a}{r_{DA}})^6\right\}$$

where $\varepsilon$ is dielectric constant, $a$ is effective Vander Waals coefficient for the interaction between neutral donor and neutral acceptor, and $r_{DA}$ is distance between neutral donor and acceptor respectively. The increased excitation intensity raises the temperature and hence causes increase in $r_{DA}$, in turn the DAP peak moves away from $FX_A^{n=1}$-2LO peak which eventually disappears at high intensity.

*Table-1 Comparison of free and bound exciton transitions in polycrystalline ZnO (present work) and single crystalline ZnO (published works).*

|  | $FX_A^{n=1}$ (eV) | $FX_A^{n=2}$ (eV) | $FX_B^{n=1}$ (eV) | $D_IX_A$ (eV) |
|---|---|---|---|---|
| Present work | 3.378 | 3.414 | 3.391 | 3.347 |
| Reynolde et-al[a] | 3.3773 | 3.4221 | 3.3895 | 3.362 |
| Teke et-al[b] | 3.3771 | 3.4206 | 3.3898 | 3.3605 |
| Hamby et-al[c] | 3.378 |  | 3.385 | 3.3605 |

a= Ref 6, b=Ref 15, c=Ref 10

Table 1 summaries our results and earlier reported results. Our measurement on bulk polycrystalline ZnO shows slight deviation from earlier reported results. The results for A and B free exciton transition energy are in accordance with previous published work, whereas, most intense peak of $D_1X_A$ is at lower energy than earlier published results. Based on correlated magnetic resonance experiment, Meyer el-al[14] suggested that neutral donor bound peak at 3.3628 eV is due to exciton bound to hydrogen and it easily annihilated by annealing at higher temperature (T> 800 $^0$C). Since, we use sintered sample at 1000 $^0$C, therefore, it is expected that donor bound exciton at 3.3628 eV is annihilated, thus not seen in main PL profile. Annealing studies also show that as the annealing temperature is increased, the higher energy PL peaks disappear and at annealing temperature of 800 $^0$C essentially, all of the emission intensity goes into the lower energy emission peak, which is due to nearest neighbor alignment. Sintering at 1000 $^0$C results in defect diffusion and produces nearest neighbor defect pairs. The PL spectrum of this defect pairs shows polarization properties. The details of PL study of defect pairs are described elsewhere.[7]

Figure 5 (a) shows the EPR spectra of sintered ZnO in temperature range of 275K-100K, a standard sample (DPPH) of known g ( 2.0036) value is used as reference to find out exact value of g. EPR line of DPPH and ZnO occurs at 3366 Gauss and 3445 Gauss respectively. Interestingly an EPR spectrum of ZnO at 100K seems very asymmetric. For clarity, it is re-plotted in figure 5(b), while inset shows the integrated EPR spectra. It has three peaks at g value of 1.985 (I), 1.956 (II) and 1.939 (III). EPR at g=1.956 and g=1.985 is due to shallow donors Zn interstitial and oxygen vacancy respectively. Central peak (II) is very asymmetric; and seems as envelop of two lines as shown with dotted line with g value at 1.959 and 1.957. Asymmetry of peak (II) indicates that two different shallow donors are present in ZnO. These donors have almost identical g values, which prohibit their clear separation. Hofman et al has resolved the two type of donor with pulsed 95 GHz EPR spectrometer.[16] They observe g= 1.9569 and 1.9571 for two donor. This is in close agreement with our experiment. Anisotropy in sintered ZnO are due to defect diffusion, $V_O$, $Zn_i$, $V_{Zn}$ or combination of all. $Zn_i$ is known to diffuse at temperature around 500 $^0$C.[38] Combination of local density approximation ( LDA) and correlation energy (U) also supports anisotropy in ZnO.[39] It has been shown that lattice relaxations around $V_0$ are large and very different in different charge states ($V^+_0$, $V^{++}_0$). For $V_0$, the four Zn nearest neighbor are displaced inward by 12% of equilibrium Zn-O bond length, whereas, for $V^+_0$ and $V^{++}_0$ the displacement are outward by 2% and 23 %. Thermal activation energy of donors can also be estimated by temperature dependence of PL peak intensity using equation 3. In this case, sample temperature was kept constant while, excitation intensity was varied from 0.5 MW-cm$^{-2}$ to 4.4 MW-cm$^{-2}$. The thermal activation energy of exciton is estimated from the theoretical fit of equation 3.

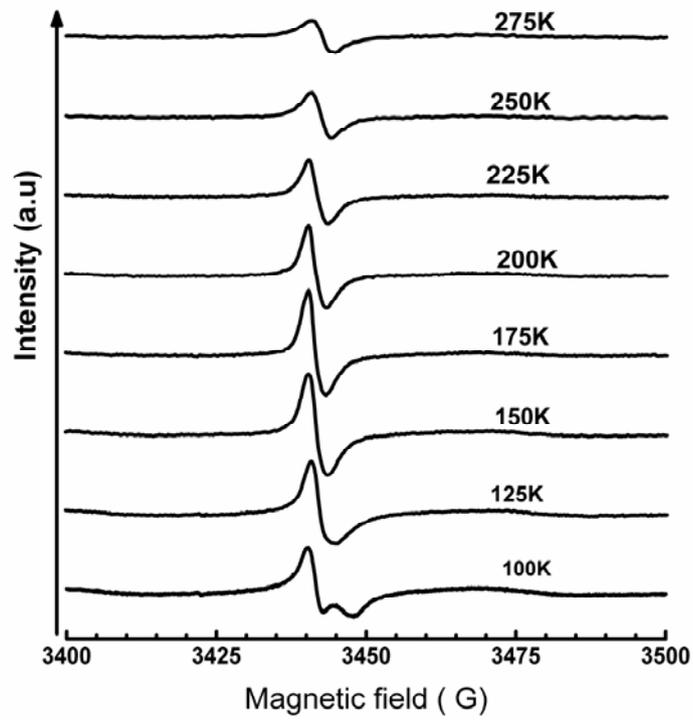

(a)

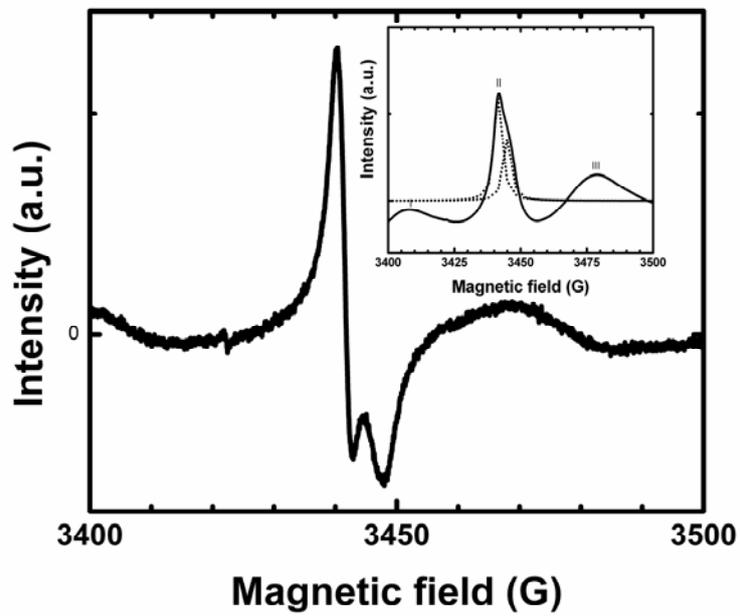

(b)

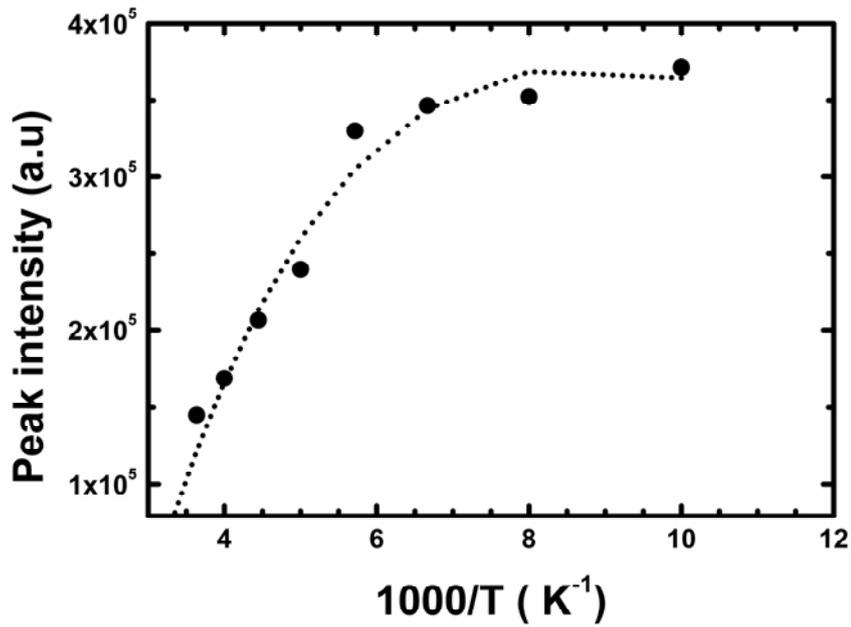

(c)

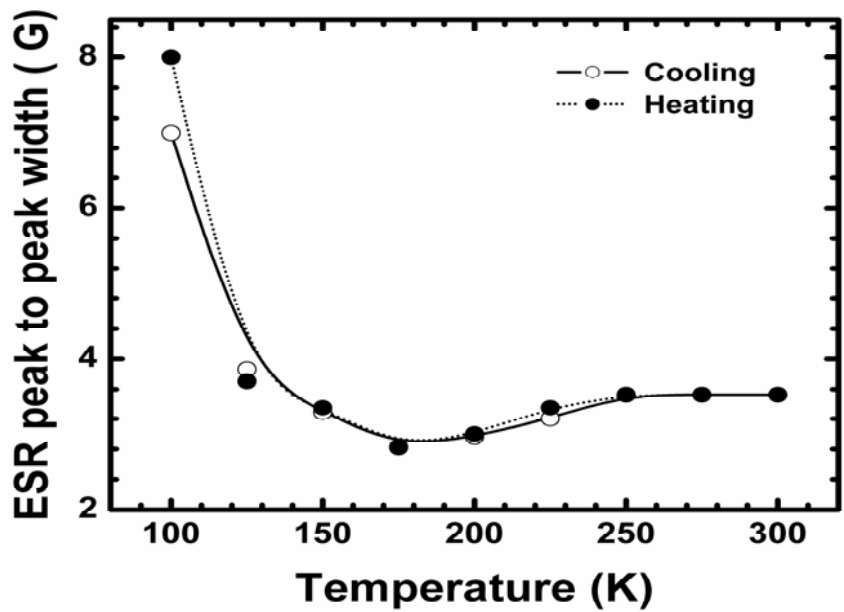

(d)

**Figure 5**. EPR spectra of sintered ZnO powder, (a) EPR profile at various temperature. (b) EPR spectra at 100K on extended scale, inset shows integration of EPR profile dotted line is Lorentzian de-convolution. (c) variation of EPR peak-to-peak intensity with temperature, solid line is theoretical fit of equation 4 to experimental data points. (d) Variation of peak-to peak width with temperature.

Figure 6 shows the thermal activation energy/binding energy of donor bound exciton, inset shows the variation of PL peak intensity with temperature at different excitation intensity, solid lines shows theoretical fit to experimental data points. Since, increase of excitation intensity causes increase of temperature within focal volume hence further affecting the lattice relaxation and defect pair rearrangement. Activation energy of donor varies from 12-31 meV with increase of excitation intensity from 1.82 MW-cm$^{-2}$ to 4.4 MW-cm$^{-2}$.

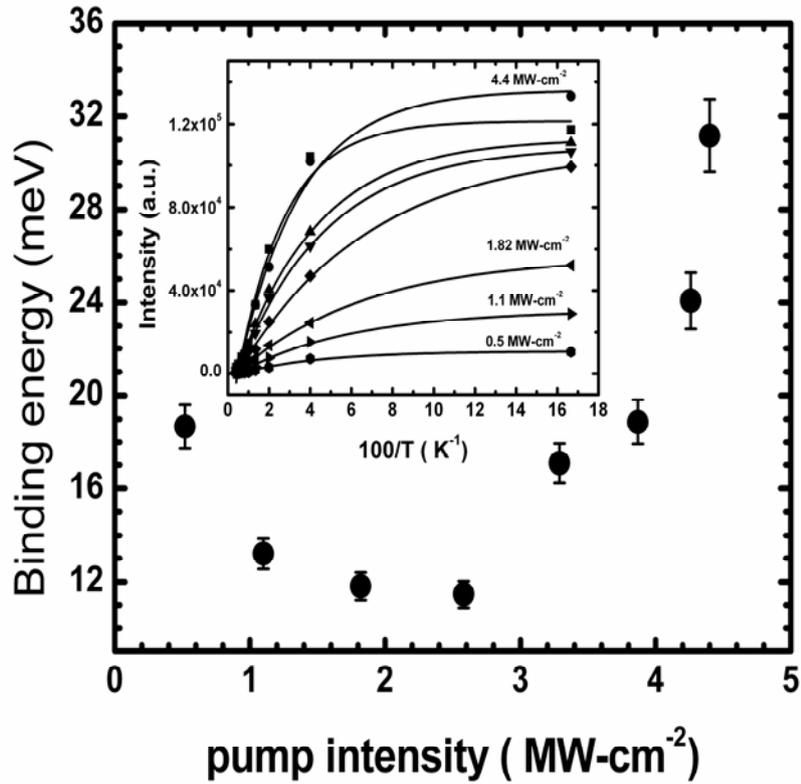

**Figure 6**. Variation of binding energy of $D_1X_A$ with excitation intensity. Inset shows variation of PL peak intensity with temperature at different excitation intensity, solid line is theoretical fit of equation 2 to experimental data points.

Further, in order to investigate whether our EPR signal corresponds to two donors as suggested by 95 GHz EPR and Hall measurement,[16] we carried out two type of analysis. (i) We investigate the temperature dependence of EPR peak intensity (ii) variation of peak-to-peak width of EPR signal with temperature.

Figure 5(c) shows the variation of EPR peak intensity with temperature. To analyze this behavior we used the following expression [16]

$$I_{EPR}(T) = \frac{I_{EPR}(0)}{1 + K_1 \exp(-\frac{E_1}{2k_B T}) + K_2 \exp(-\frac{E_2}{2k_B T})} \quad 4$$

Where, $E_i$ (i=1,2) is the thermal activation of the process i with pre factor $K_i$. It is evident that EPR peak intensity decreases with increase of temperature. This is because at low temperature both donors are in paramagnetic state. When temperature is increased donors electrons are released in conduction band leaving them in positive diamagnetic charged state. Dotted line is the theoretical fit of equation 4 to the experimental data points. This yields $E_1$=18 meV ( T~216 K )and $E_2$= 44.6 meV ( T~530 K). Hofmann,[16] also show two type of donor in broad temperature range (40K-300K) with thermal activation energy in range of 35 meV and 66 meV. Figure 5(d) show variation of peak-to –peak line width (δ) defined as the width between points of maximum signal (positive and negative). δ decreases from 3.53 Gauss to 2.82 Gauss with decrease in temperature from 300K to 200K. Further, δ increase from 2.82 Gauss to 7 Gauss with decrease in temperature from 200K to 100K. The increase of δ below 200K indicates strong spin-orbit and/or dipolar interaction. Strength of crystal field splitting (CFS) and spin orbit coupling (SOC) in sample was calculated using PL data. CFS and SOC is calculated using formula [30]

$$\left.\begin{array}{c}CFS\\SOC\end{array}\right\} = \frac{1}{2}\left[\Delta_{CB} - \Delta_{BA} \pm \sqrt{2\Delta_{CA}^2 - \Delta_{BA}^2 - \Delta_{CB}^2}\right] \quad 5$$

where $\Delta_{CB}$= $E_g(C)-E_g(B)$, $\Delta_{BA}$= $E_g(B)-E_g(A)$, $\Delta_{CA}$= $E_g(C)-E_g(A)$, $E_g(A)$=3.378 eV, $E_g(B)$= 3.391 eV and $E_g(C)$= 3.435 eV from reference 15. This calculation yields CFS= 42 meV and SOC= -20 meV. CFS is in accordance with previous reported value of 31 meV.

## IV CONCLUSION

In conclusion, photoluminescence profile of polycrystalline bulk ZnO at 6 K is dominated by donor bound exciton peak with lower energy, while other free exciton transitions are in accordance with PL profile of single crystalline ZnO. EPR result at 100K shows that sintering at high temperature causes re-arrangement of defect structure and form new defect pairs. It is worth mentioning that the observed behavior is due to in homogeneous strain and lattice relaxation around oxygen vacancy. Dominance of 1LO peak at room temperature is due to shift of DBE PL peak toward lower energy side and 1LO peak towards higher energy side with increase in pump intensity.

## ACKNOWLEDGEMENT
Author, would like to thank professor Ranjan Das, Department of Chemical Sciences, Tata Institute of Fundamental Research Mumbai India, for providing facility for Low temperature EPR measurement. We also thank Mr. Vinayak Rane and Farman Ali for useful discussion.